\documentclass[graybox]{svmult}

\usepackage{mathptmx}       
\usepackage{helvet}         
\usepackage{courier}        
\usepackage{type1cm}        
\usepackage{makeidx}         
\usepackage{graphicx}        
\usepackage{multicol}        
\usepackage[bottom]{footmisc}

\makeindex             

\begin{document}

\title*{Structure and dynamics of penumbral filaments}
\author{B.\ Ruiz Cobo, \and L.R.\ Bellot Rubio}
\institute{B.\ Ruiz Cobo, \at Instituto de Astrof\'\i sica de Canarias, Tenerife, Spain \email{brc@iac.es}
\and L.R.\ Bellot Rubio, \at Instituto de Astrof\'\i sica de Andaluc\'\i a (CSIC), 
Granada, Spain \email{lbellot@iaa.es}}
%
%
\maketitle

\vspace{-2cm}
\abstract{High-resolution observations of sunspots have revealed the
existence of dark cores inside the bright filaments of the
penumbra. Here we present the stationary solution of the heat transfer
equation in a stratified penumbra consisting of nearly horizontal
magnetic flux tubes embedded in a stronger and more vertical
field. The tubes and the external medium are in horizontal mechanical
equilibrium. This model produces bright filaments with dark cores as a
consequence of the higher density of the plasma inside the flux tube,
which shifts the surface of optical depth unity toward higher (cooler)
layers. Our results suggest that the surplus brightness of the
penumbra is a natural consequence of the Evershed flow, and that
magnetic flux tubes about 250 km in diameter can explain the
morphology of sunspot penumbra.}

\section{Introduction}
\label{sec:1}
A good knowledge of the structure of the penumbra is important to
understand various physical processes occurring in sunspots as, for
example, the Evershed flow, the brightness of the penumbra, and the
coexistence of magnetic fields with different strengths and
inclinations. Despite significant advances in the characterization of
the penumbra during the last years~\cite{Solanki2003, BellotRubio2004,
BellotRubio2007}, its basic building blocks have not been identified
unambiguously as yet.

Recent images taken with the Swedish 1-m Solar Telescope have
demonstrated that many penumbral filaments possess internal structure
in the form of a dark core \cite{Scharmeretal2002,
RouppevanderVoortetal2004, Sutterlinetal2004}. The central obscuration
is surrounded by two narrow lateral brightenings, both of which are
observed to move with the same speed and direction as a single entity.

The origin of these structures remains enigmatic. In a recent paper,
Spruit \& Scharmer~\cite{SpruitScharmer2005} examined the idea that
dark-cored filaments could be the signature of field-free gaps just
below the visible surface of the penumbra. They suggested that the top
of the gaps would be observed as dark cores due to their increased
plasma density that shifts the $\tau=1$ level to higher (cooler)
layers. They also speculated that the large variations of the magnetic
field strength and inclination around and above the gaps could explain
the net circular polarization of spectral lines observed in the
penumbra. For the moment, no radiative transfer calculations have been
performed to sustantiate these claims.

On the other hand, the idea that a dark-cored penumbral filament
represents a magnetic flux tube embedded in a stronger and more
vertical ambient field is very appealing. This model of the penumbra,
known as the uncombed model~\cite{SolankiMontavon1993}, has been used
with great success to reproduce the polarization profiles of visible
and infrared spectral lines emerging from sunspots. Numerical
simulations of the evolution of penumbral flux tubes have been carried
out by, e.g., Schlichenmaier~\cite{Schlichenmaieretal1998}. The
simulations offer an explanation for the existence of the Evershed
flow and explain, at least qualitatively, the most important features of the
penumbra. Given the success of the uncombed model, it seems natural to
examine whether it can also explain the existence of dark-cored
filaments.

Schlichenmaier et al.~\cite{Schlichenmaieretal1999} studied the radiative cooling of
hot penumbral flux tubes surrounded by a non-stratified, initially isothermal
atmosphere. They found that the tubes cool down very quickly in the absence of
energy sources, reaching thermal equilibrium with the external medium in only
a few tens of seconds.  No dark cores would be observed under these conditions.

Here we extend the work of Schlichenmaier et al.~\cite{Schlichenmaieretal1999} by solving the
stationary heat transfer equation in a 2D stratified penumbra. We model
penumbral filaments as magnetic flux tubes embedded in a stronger and more
vertical ambient field. The two sources of energy we consider are Ohmic
dissipation and a hot Evershed flow. Our calculations show that the
temperature distribution is not symmetric around the tube axis, and that one
such tube would be observed as a dark-cored filament due to the higher density
(hence higher opacity) of the plasma within the tube.  Interestingly, the
observed intensity contrast of the dark core relative to the surroundings 
is not well reproduced without the Evershed flow. More details about the
equations, numerical methods, results, and implications can be found in 
\cite{ruizcobo+bellotrubio2008}.

\section{The model}
\label{sec:2}
We describe a bright penumbral filament as a cylindrical magnetic flux
tube of radius $R$ embedded in a stratified background atmosphere. 
Spectropolarimetric observations suggest that the inclination
and strength of the magnetic field are different in the tube
and the external medium.  
This produces electrical currents at the interface. We account for
them assuming that the tube and the surrounding atmosphere are
separated by a current sheet of thickness $\delta$ which extends 
from $r= R - \delta$ to $r=R$. 

\begin{figure}
\begin{center}
\includegraphics[bb=35 5 485 160,clip,scale=.82]{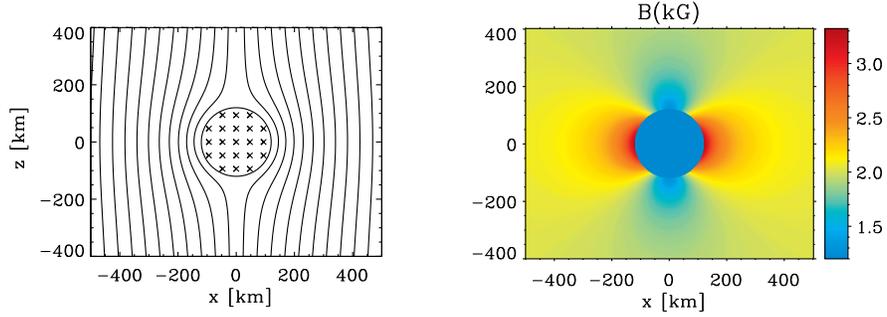}
\end{center}
\caption{{\it {Left:}} Magnetic field lines in the plane perpendicular to the tube's
axis (the $xz$ plane). The circle centered at $z=0$ represents the 
tube's boundary. {\it {Right:}} Field stregth distribution.}
\label{fig:1}
\end{figure}

We require the tube and the surroundings to be in lateral mechanical
equilibrium, i.e., the total (gas plus magnetic) pressures inside and
outside the tube must be the same at the same geometrical height
$z$. The ideal gas law then determines the gas density if the
temperature and the magnetic field strength are known. The temperature
distribution is obtained by solving the stationary heat transfer
equation. Since the only purpose of the present calculations is to
identify the physical mechanism responsible for the existence of dark
cores, we adopt the simplest magnetic configuration possible, namely a
potential field. The field is determined from the conditions $\nabla
\cdot \vec{B} =0$ and $\nabla \times \vec{B} = 0$ in the $xz$-plane,
i.e. we neglect variations along the tube axis because they are much
smaller than variations perpendicular to it~\cite{Borreroetal2004}.
In the tube's interior, the field is assumed to be homogeneous and
directed along its axis. Figure~\ref{fig:1} shows the magnetic
configuration of the model with $R=120$~km, $B_{\rm t}= 1200$~G,
$\gamma_{\rm t}=90^\circ$, $B_{\rm b}=2000$~G, and $\gamma_{\rm
b}=40^\circ$. Note that the field lines in the background wrap around
the tube, leading to enhanced field strengths on
either side of the tube and a reduction of the field right above and
below it. Note also that the field component normal to the
tube's boundary vanishes, implying that the inclination of the
external field does not remain constant.  In particular, immediately
above the tube the background field is more horizontal than
$\gamma_{\rm b}$.

\section{Heat transfer equation}

We solve the stationary heat transfer equation, $\nabla \cdot \vec{F}
= S,$ where $\vec{F} = \vec{F}_{\rm r} + \vec{F}_{\rm c}$ is the total
energy flux (with $\vec{F}_{\rm r}$ the radiative flux and
$\vec{F}_{\rm c}$ the convective flux), and $S$ represents the various
energy sources, including Joule heating and heat deposited by the
Evershed flow. The radiative flux is computed using the diffusion
approximation (e.g., \cite{Mihalas1978}), i.e., $\vec{F}_{\rm r}
= - k_{\rm r} \, \nabla T$, with $k_{\rm r}$ the radiative thermal
conductivity and $T$ the temperature. Following
\cite{Schlichenmaieretal1999}, we take $k_{\rm r} = 16 \, D_{\rm F} \,
\sigma \, T^3/(\kappa_{\rm R} \rho)$, where $\sigma$ is
Stefan-Boltzmann constant, $\kappa_{\rm R}$ the Rosseland mean
opacity, $\rho$ the plasma density, and $D_{\rm F}$ the flux limiter
originally introduced by Levermore \&
Pomraning~\cite{LevermorePomraning1981}. For the calculation of the
convective flux we adopt a linearized mixing length
approach~\cite{Moreno-Insertisetal2002} and use $\vec{F}_{\rm c}=
-k_{\rm c} \, [\nabla T - ({\rm d}T/{\rm d}z)_{\rm ad}]$. The quantity
$({\rm d}T/{\rm d}z)_{\rm ad} = T ({\rm d}P/{\rm d}z) \nabla_{\rm ad}$
represents the adiabatic temperature gradient and $\nabla_{\rm ad}$
the double-logarithm isentropic temperature gradient. The convective
transport coefficient $\kappa_{\rm c}$ is calculated using the
linearized approximation of Spruit~\cite{Spruit1977}.

\begin{figure}[t!]
\sidecaption
\includegraphics[scale=.82]{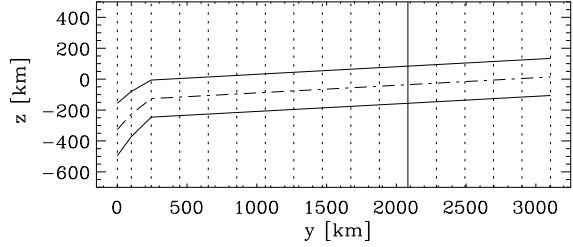}
\caption{Cut of the computational domain at x=0 km, showing a tube 120 km in radius. 
The vertical lines mark the 17 $xz$-planes where we solve the heat transfer 
equation. The line at $y$=2083 km indicates the position of the $xz$-cut displayed in Fig.\ 3 }
\label{fig:2}       
\end{figure}

The heat transfer equation is solved numerically with Neumann boundary
conditions given by $\delta T(x,z=0) = 0 \,\, \forall x $ and
$\dot{\delta T}(x=0,z) =\dot{\delta T}(x=L,z) = 0 \,\, \forall
z$. Here, $\delta T$ stands for temperature departures from the
solution without tube. The second condition forces the temperature
perturbations to be minimum at the lateral boundaries of the
computational domain.

\subsection{Simulation setup}
The computational box is a cube extending 1000 km in the
$x$-direction, 3100 km in the $y$-direction, and 1200 km in the
$z$-direction.  Initially, the cube is filled with an unperturbed
atmosphere (the cool model of Collados et al.~\cite{Colladosetal2004})
with magnetic field strength $B_{\rm b}=2000$~G and field inclination
$\gamma_{\rm b} = 40^\circ$.  A penumbral flux tube is embedded in
this atmosphere. The tube has a radius $R= 120$ km, a current sheet 2
km thick, a field strength $B_{\rm t}= 1200$~G, and magnetic field
inclinations varying between 45$^\circ$ and 87$^\circ$ as shown in
Fig.~\ref{fig:2}. The axis of the tube is placed at a height
determined by the field inclination and the distance along the tube,
starting with $z=-326$~km at $y=0$~km.

The 2D stationary heat transfer equation has been solved in 17 $xz$-planes at
different $y$ values (cf.\ Fig~\ref{fig:2}). Each plane is discretized in
$501 \times 601$ grid points, the step size being 2 km. 

\section{Results and discussion}
\label{evershed_energy}
The Evershed flow is a radial, nearly horizontal mass outflow observed
in sunspot's penumbrae. It is associated with the more inclined
magnetic field component of the penumbra~\cite{Titleetal1992,
Stanchfieldetal1997, WestendorpPlazaetal2001, BellotRubio2004,
Rimmele2008} and often exhibits supersonic velocities \cite{Bellotetal2004}. The
Evershed flow has been modeled as a material flow along thin flux tubes \cite{Schlichenmaieretal1998}.

An Evershed flow of hot plasma along a magnetic flux tube with velocity
$\vec{v}_{\rm E}$ produces a flux of energy $\vec{F}_{\rm E}$ whose divergence
can be evaluated from the entropy equation as
\begin{equation}
\nabla \cdot \vec{F}_{\rm E} = \rho c_V \vec{v}_{\rm E} \, [\nabla T - ({\rm
d}T/{\rm d}z)_{\rm ad} \vec{e}_{\rm z}], 
\label{flujoevershed}
\end{equation}
where $c_V$ is the specific heat at constant volume. To account for
the heating induced by the Evershed flow, we set $S=j^2/\sigma +
\nabla \cdot \vec{F}_{\rm E}$ in the stationary heat transfer equation. We
assume that the Evershed flow is parallel to the magnetic field inside
the tube (\cite{Bellotetal2004}), and that its velocity changes
with radial distance as dictated by mass conservation due to the
decrease of the density with height.

\begin{figure}[t!]
{\includegraphics[scale=.28]{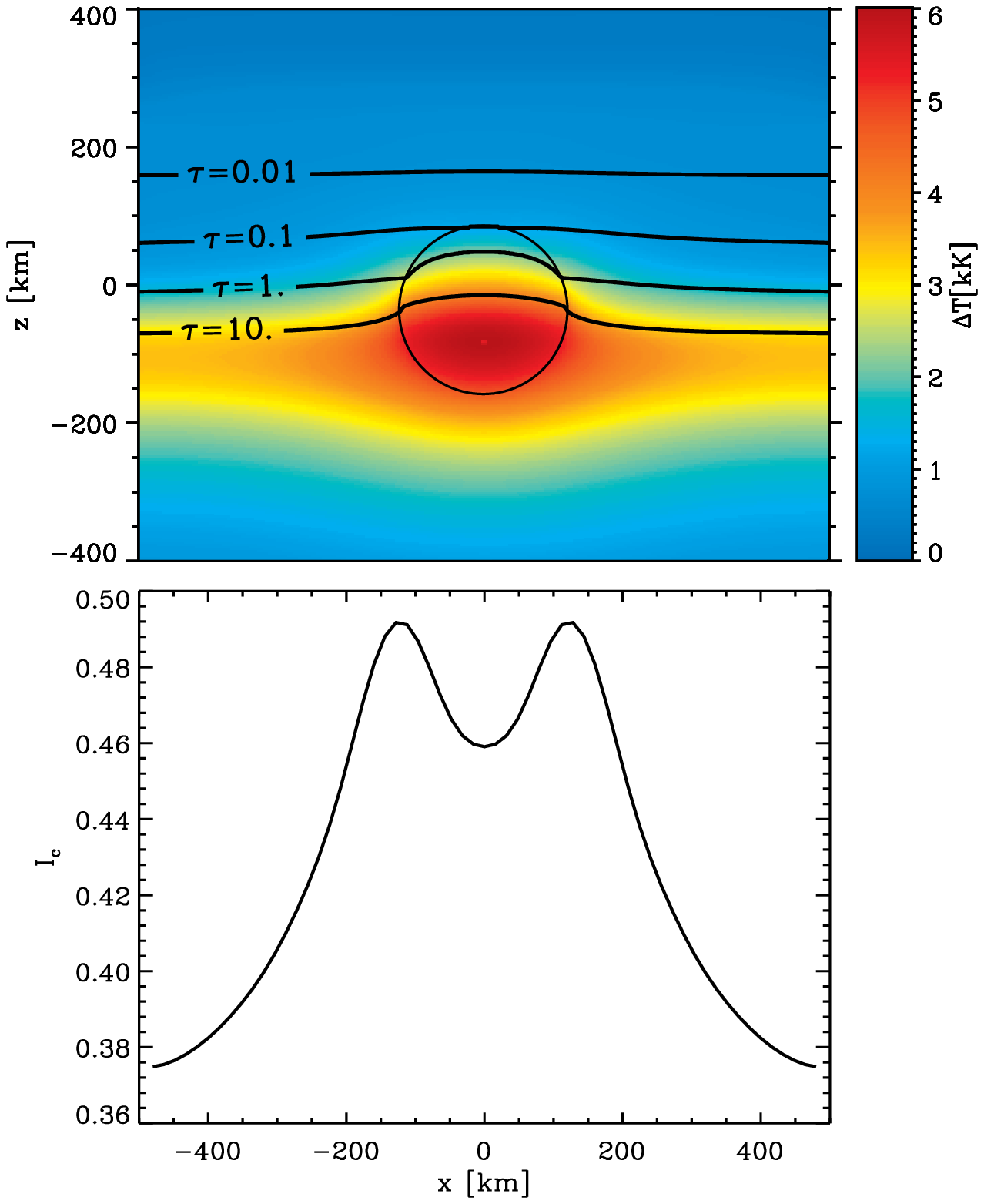}}{\includegraphics[scale=.8]{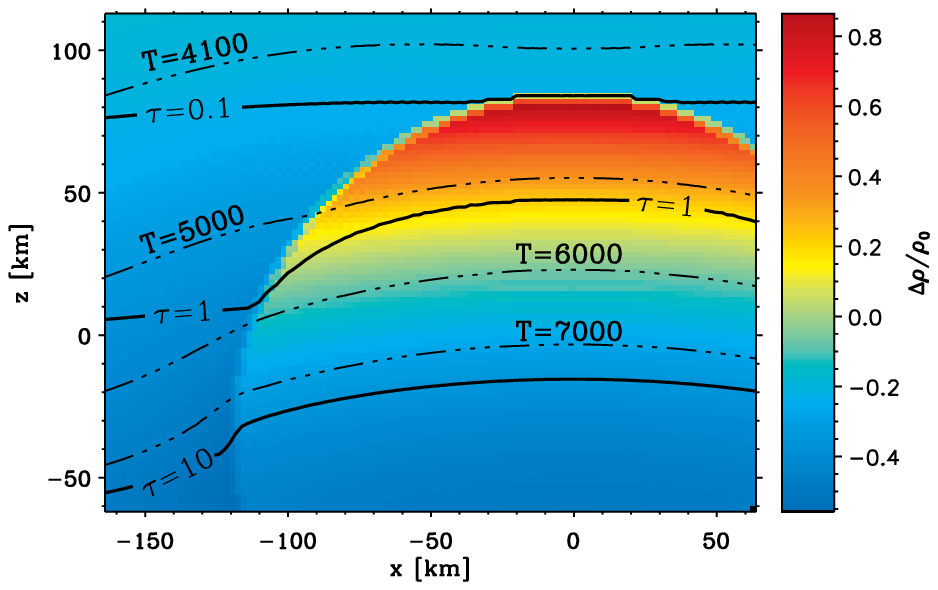}}
\caption {Top left panel: Temperature perturbation $\delta T$ in the $xz$--plane at $y=2083$~km.
The sources of energy are Ohmic dissipation and a hot Evershed
flow. The circle represents the flux tube. Solid lines are lines of
constant Rosseland optical depth. Bottom left panel: continuum
intensity at 487.8 nm emerging from the $y=2083$~km plane, convolved
with the Airy point-spread function of a 1-m telescope. The values are
normalized to the quiet Sun continuum intensity. Right panel: Section
of the same plane showing gas density perturbations
$\Delta\rho/\rho_0$ induced by the weaker field of the tube and the
Evershed flow. Dash-dotted lines represent isotherms. Solid lines
indicate lines of constant Rosseland optical depth.}
\label{fig:4}       
\end{figure}

Figure~\ref{fig:4} shows the results of the calculations for the
$xz$-plane at $y=2083$~km, assuming $v_{\rm E}$ to be 7~km~s$^{-1}$
at $y=0$~km. As can be seen in the upper left panel, the Evershed flow
causes an intense heating of the tube and the surroundings. The right
panel demonstrates that the $\tau_{\rm R}=1$ level is shifted upwards within
the flux tube. The reason is the weaker field of the tube, which increases
its density and opacity. The $\tau_{\rm R}=1$ level then moves to higher 
layers, where the temperature is lower. This produces a dark-cored 
penumbral filament (cf.~the lower left panel of Fig.~\ref{fig:4}).

\begin{figure}[t!]
\begin{center}
\includegraphics[bb=0 15 423 96,clip,scale=.72]{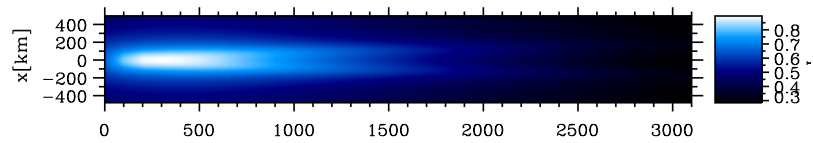}
\end{center}
\caption{Continuum image of the flux tube at 487.8 nm. The Evershed flow has a 
velocity of 7~km~s$^{-1}$ and induces a temperature excess of 7500 K
within the tube at y=0 km.}
\label{fig:6}    
\end{figure}

Figure~\ref{fig:6} displays a continuum image of the flux tube at 487.8
nm as it would be observed through a 1-m telescope at disk center ($\mu = 1$).
This image has been obtained solving the radiative transfer equation for all
$xz$-planes of the computational box. Note that the observed characteristics
of dark-cored penumbral filaments are qualitatively well reproduced: the
central obscuration is flanked by two lateral brightenings, the intensity 
of the dark core is about 85\% that of the lateral brightenings, and the whole 
filament is significantly brighter than its surroundings~\cite{Scharmeretal2002, Sutterlinetal2004, RouppevanderVoortetal2004}. 

The calculations presented here strongly support the view that the
penumbra is formed by small (but thick) magnetic flux tubes, as inferred from the
analysis of high resolution filtergrams and spectropolarimetric
measurements~\cite{SolankiMontavon1993, Solanki2003, BellotRubio2004}.

\begin{acknowledgement}
This work has been supported by the Spanish Ministerio de Ciencia y Tecnolog\'{\i}a under projects AYA2001-1649, ESP2003-07735-C04-03, and {\em Programa Ram\'on y Cajal}
\end{acknowledgement}

%
%
%

\end{document}